\documentclass[]{aa}  

\usepackage{graphicx}
\usepackage{txfonts}
\usepackage{epsfig,graphics,graphicx,bm,amssymb}
\usepackage[normalem]{ulem}

\usepackage[dvipsnames]{xcolor}

\usepackage[]{natbib}
\usepackage[hidelinks]{hyperref}

\begin{document}

   \title{3D stellar motion in the axisymmetric Galactic potential and the \lowercase{\textit{\Large e}}--\lowercase{\textit{\Large z}} resonances}

   \author{Tatiana A. Michtchenko
          \inst{1}
          \and
          Douglas A. Barros\inst{2}
          }

   \institute{Universidade de S\~ao Paulo, IAG, Rua do Mat\~ao, 1226, Cidade Universit\'aria, 05508-090 S\~ao Paulo, Brazil\\
              \email{tatiana.michtchenko@iag.usp.br}
         \and
             Rua Sessenta e Tr\^es, 125, Olinda, 53090-393 Pernambuco, Brazil\\
             \email{douglas.barros@alumni.usp.br}
             }

   \date{Received ---- --, ----; accepted ---- --, ----}

 
  \abstract
   {The full phase-space information on the kinematics of a huge number of stars provided by the Gaia Data Release 3 raises the demand for a better understanding of the 3D stellar dynamics.}
   {In this paper, we investigate the possible regimes of motion of stars in the axisymmetric approximation of the Galactic potential, applying a 3D observation-based model {developed elsewhere}. The model consists of three components: the axisymmetric disk, the central spheroidal bulge and the spherical halo of dark matter. The axisymmetric disk model is divided into thin and thick stellar disks and \textrm{H}\,{\scriptsize I} and $\mathrm{H}_2$ gaseous disks subcomponents, by combining three Miyamoto-Nagai disk proﬁles of any model order (1, 2, or 3) for each disk subcomponent,  to reproduce a radially exponential mass distribution. The physical and structural parameters of the Galaxy components are adjusted by observational kinematic constraints.}
   {The phase space of the two-degrees-of-freedom model is studied by means of the Poincaré and dynamical mapping, the dynamical spectrum method and the direct numerical integrations of the Hamiltonian equations of motion.}
   {For the chosen physical parameters, the nearly-circular (close to the rotation curve) and low-altitude stellar behaviour is composed of two  weakly coupled simple oscillations, radial and vertical motions. The amplitudes of the vertical oscillations of these orbits are gradually increasing with the growing Galactocentric distances, in concordance with the exponential mass decay assumed. However, for increasing planar eccentricities $e$ and the altitudes over the equatorial disk $z$, new regimes of stellar motion emerge as a result of the beating between the radial and vertical oscillation frequencies, which we refer to as $e$--$z$ {\it resonances}. The corresponding resonant motion produces characteristic sudden increase/decrease of the amplitude of the vertical oscillation, bifurcations in the dynamical spectra and the chains of islands of stable motion in the phase-space.}
   {The results obtained can be useful in the understanding and interpretation of the features observed in the stellar 3D distribution around the Sun.}


   \titlerunning{3D stellar motion and the $e$--$z$ resonances}
   
   \authorrunning{T. Michtchenko \& D. Barros}
   
   \maketitle
%

\section{Introduction}
\label{sec:intro}

In the past decades, there has been a growing set of evidence that the disk stars of the Milky Way galaxy exhibit density and velocity structures in several phase-space planes of the Galactic disk, in both the radial and vertical directions. The long-known stellar warp in the outer disk, a vertically asymmetric distribution of stars close to the Galactic plane, is seen as a bending of the plane upwards in the first and second Galactic quadrants (longitudes $0^{\circ}\leq l \leq 180^{\circ}$) and downwards in the third and fourth quadrants (longitudes $180^{\circ}\leq l \leq 360^{\circ}$) \citep{Momany2006}. {Recently, a number of stellar density substructures, appearing as overdensities or dips, have been found in the LAMOST data, at several radial and vertical positions in the Galactic disk \citep{Wang2018}.}

Large-scale vertical motions as a Galactic North--South asymmetry have been revealed by spectroscopic and astrometric surveys such as SEGUE \citep{Widrow2012}, RAVE \citep{Williams2013}, LAMOST \citep{Carlin2013}, as well as the second \textit{Gaia} data release \citep{Katz2018Gaia}. Such bulk vertical motion of stars present a wave-like behaviour, which has been referred to as bending and breathing motions, with stars coherently moving towards or away from the Galactic mid-plane, on both sides of it \citep{Kawata2018,Ghosh2022,Khachaturyants2022}. Several dynamical processes have been evoked to explain these non-zero vertical motions: the ones from external origins such as the passing of the Saggitarius dwarf galaxy through the Milky Way disk \citep{Gomez2013}, or the excitation of the disk due to interactions with dark matter subhaloes \citep{FeldmannSpolyar2015}; and the ones from non-axisymmetric internal perturbations such as the breathing motion induced by the Galactic bar \citep{Monari2015}, or by the action of the spiral density waves \citep[among others]{Faure2014,Ghosh2022,Khachaturyants2022}.

The density structures present in the $R$--$V_\varphi$ plane (Galactocentric radius versus azimuthal velocity) of disk stars, known as diagonal ridges, clearly visible in the distribution of the mean Galactic radial velocity $\left<V_R\right>$, can also be seen in the vertical direction from maps of the mean absolute distance from the disk mid-plane $\left<|z|\right>$ and the mean vertical velocity $\left<V_z\right>$ of the stellar distribution \citep{Khanna2019,Wang2020}. This attribute may be indicative of a coupling between planar and vertical motions of the stars in the disk. Several attempts to unravel the origin of these structures have been presented in the literature, with some of them relying on simulations of external mechanisms like the Sagittarius dwarf galaxy perturbation \citep{Antoja2018,Laporte2019,Khanna2019}, and others from simulations that take into account internal dynamics such as the bar or the spiral resonances \citep{Hunt2018,Michtchenko2018,Fragkoudi2019,Barros2020}.

Another striking feature associated with the vertical motion of the stars in the Galactic disk is the phase-space spiral (or the so-called snail shell) present in the $z$--$V_z$ plane. Many authors interpret it as evidence of an ongoing phase mixing in the vertical direction of the disk due to an influence of an outer perturbation \citep{Antoja2018,BinneySchonrich2018,BlandHawthorn2019,Laporte2019}, or caused by the instability of a buckling bar \citep{Khoperskov2019}, or even due to the earlier discussed vertical bending waves \citep{DarlingWidrow2019} or vertical breathing motion \citep[][for two-armed phase spirals]{Hunt2022} in the disk. Alternatively, \cite{Michtchenko2019} showed that the phase spiral can be well-comprehended by the dynamical effects of the stellar moving groups in the solar neighbourhood.

We suggest to analyse the above-cited Galactic structures and investigate their causes by studying the stellar dynamics in the following steps:
\begin{itemize}
\item Modeling a 3D Galactic gravitational potential. In this work, we use the \cite{Barros2016} axisymmetric Galactic potential model, adjusted to recent observational data. Firstly, the chosen potential is suitable to the study of the vertical stellar motion and, secondly, it is simple in analytical forms that gives a good representation of the radially exponential mass distribution of the Galaxy. Moreover, any non-axisymmetric mass effects (e.g., due to the central bar and/or spiral arms) can be easily added to the model posteriorly;
\item Studying the Galactic model in the whole phase space, in order to obtain all the possible regimes of stellar motion that could provide reasonable explanations to the observed Galactic phase-space structures. This is, in part, the main objective of the present paper;
\item Performing numerical simulations through numerical integrations of stellar orbits, to verify whether our achievements  are satisfactory. This is a topic for a future work.
\end{itemize}

In the present paper, we focus on the study of the stellar orbits in the Galactic disk assuming a zeroth-order approximation of a time-independent and axisymmetric Galactic gravitational potential. The whole phase space around the Sun is investigated through techniques widely used in the Celestial Mechanics. Resonances between the radial and vertical independent modes of the stellar motion are characterized in terms of resonant zones, formed by islands of stability that capture and trap stars inside them, enhancing the stellar density, and separatrices that deplete the objects close to these regions. Our goal, for a subsequent work, is to investigate possible associations between the observed Galactic phase-space structures and the resonant zones originated from the commensurabilities between the radial and vertical frequencies of the stellar motion.

This paper is organized as follows. In Section 2, we describe the 3D axisymmetric Galactic gravitational potential model used for the construction of the Hamiltonian function and the calculus of the stellar orbits. In Section 3, we study the 3D stellar dynamics on the representative plane $R$--$V_\varphi$ in the radial and vertical directions. In Section 4, regimes of stellar motion are analysed in terms of the beating between the planar and vertical frequencies of oscillation, and, in Section 5, we extend the analysis for a wide range of initial vertical velocities $V_z$. Concluding remarks are drawn in the closing Section 6.


\section{3D Galactic model}
\label{sec:model}

\begin{table*}
\caption{Physical and structural parameters of the disk components of the axisymmetric Galactic model}             
\label{tab-init}      
\centering   
\begin{tabular}{l c c c c c c c}  
\hline\hline       
Disk component & $M_1$            & $a_1$& $M_2$            & $a_2$& $M_3$            & $a_3$& $b$ \\ 
          &($10^{10}M_\odot$)&(kpc) &($10^{10}M_\odot$)&(kpc) &($10^{10}M_\odot$)&(kpc) & (kpc) \\  
\hline                    
Thin disk   & 2.282& 3.859& 2.342& 9.052& -1.846& 3.107& 0.243\\
Thick disk  & 0.061& 0.993& 4.080& 6.555& -3.521& 7.651& 0.776\\
\textrm{H}\,{\scriptsize I} & 2.217& 9.021& 2.350& 9.143& -3.303& 7.758& 0.168\\
$\mathrm{H}_2$ & 1.005& 6.062& 0.177& 3.141& -0.907& 4.485& 0.128\\
\hline                  
\end{tabular}
\end{table*}

\begin{table}
\caption{Physical and structural parameters of the spheroidal components of the Galactic model}
\label{tab-init-2}      
\centering   
\begin{tabular}{l c c}
\hline\hline 
Bulge & $M_b$              & $a_b$  \\
      & ($10^{10}M_\odot$) & (kpc)  \\
\hline
  & 2.54   & 0.425  \\
\hline\hline
Dark halo & $r_h$ & $v_h$   \\
          & (kpc) & (km\,s$^{-1}$)   \\
\hline
          & 5.56  & 169.77 \\  
\hline
\end{tabular}
\end{table}

\begin{figure}
	\includegraphics[width=1.0\columnwidth]{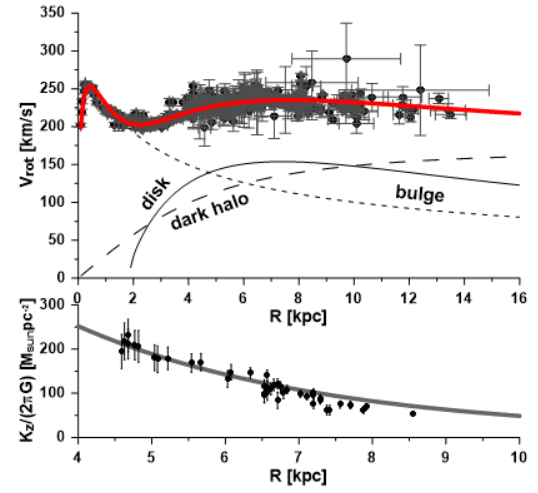}
    \caption{Top: Rotation curve of the Galaxy. The observed rotation curve is represented by the points with error bars, which indicate masers data from high-mass star-forming regions \citep{Reid_etal2019, Rastorguev_etal2017}, and \textrm{H}\,{\scriptsize I} and CO tangent-point data \citep{Burton_Gordon1978, Clemens1985, Fich_Blitz_Stark1989}. The red curve shows the analytical rotation curve expressed by Eq.\,(\ref{eq:rot-curve}). The contributions of the modelled disk (continuous curve), bulge (short-dashed line) and dark halo (long-dashed line) to the axisymmetric Galactic potential are also shown.
    {Bottom: Vertical force $K_z$ at $z=1.1$\,kpc as a function of the Galactic radius. The black dots with error bars are from observation measurements by \cite{Bovy_Rix2013}, while the grey solid curve shows the $K_z$ force radial profile at $z=1.1$\,kpc resulting from our Galactic model.}
    }
\label{fig:rot-curve}
\end{figure}

Here, we briefly describe the 3D model for the axisymmetric Galactic potential and refer the reader to \cite{Barros2016} for more details. The model considers the contributions of three Galactic components into the potential; they are the axisymmetric disk, the central spheroidal bulge and the spherical halo of dark matter. 

The  axisymmetric disk is composed of the stellar (thin and thick disks) and gaseous (\textrm{H}\,{\scriptsize I} and $\mathrm{H}_2$ disks) components. Each of these components is a superposition of three Miyamoto–Nagai disks \citep[MN,][]{Miyamoto_Nagai1975}, which reproduces a radially exponential mass distribution \citep{Smith2015}. We adopt the 1st--, 2nd-- and 3rd--order expressions for the potential of the MN--disks at the position $R$ and $z$ (cylindrical coordinates), respectively:
\begin{equation}
\label{eq:Phi_MN1}
\Phi^1_{\mathrm{MN}}(R,z)=\frac{-GM}{\sqrt{R^{2}+(a+\zeta)^{2}}}\,,
\end{equation}
\begin{equation}
\label{eq:Phi_MN2}
\Phi^2_\mathrm{MN}(R,z)=\Phi^1_\mathrm{MN}(R,z)-\frac{GM\,a(a+\zeta)}{\left[R^{2}+\left(a+\zeta\right)^{2}\right]^{3/2}}\,,
\end{equation}
\begin{equation}
\label{eq:Phi_MN3}
\Phi^3_\mathrm{MN}(R,z)=\Phi^2_\mathrm{MN}(R,z)+\frac{GM}{3}\times\frac{a^{2}\left[R^{2}-2(a+\zeta)^{2}\right]}{\left[R^{2}+\left(a+\zeta\right)^{2}\right]^{5/2}}\,,
\end{equation}
where $\zeta=\sqrt{z^2+b^2}$, with $b$ being the vertical scale length, while $M$ and $a$ are the mass and the radial scale length of each disk component, respectively.

The potential of the thin disk is then written as a combination of three third-order MN--disks

\begin{equation}
\label{eq:Phi_thin}
\Phi_\mathrm{thin}(R,z)=\sum_{i=1}^3 {\Phi^3_\mathrm{MN}}_i(R,z)\,,
\end{equation}
while the potential of the thick disk is a combination of three first-order MN--disks
\begin{equation}
\label{eq:Phi_thick}
\Phi_\mathrm{thick}(R,z)=\sum_{i=1}^3 {\Phi^1_\mathrm{MN}}_i(R,z)\,.
\end{equation}

The contributions of the gaseous disks components \textrm{H}\,{\scriptsize I} and $\mathrm{H}_2$ into the total axisymmetric potential are, respectively,
\begin{eqnarray}
\label{eq:Phi_HI}
\Phi_\mathrm{HI}(R,z) &=&\sum_{i=1}^3 {\Phi^2_\mathrm{MN}}_i(R,z)\,,\\
\Phi_\mathrm{H_2}(R,z)&=&\sum_{i=1}^3 {\Phi^3_\mathrm{MN}}_i(R,z)\,.
\end{eqnarray}

The potential of the Galactic bulge 
is derived by a Hernquist density distribution proﬁle, in the form \citep{Hernquist1990}: 
\begin{equation}
\label{eq:Phi_bulge}
\Phi_{\mathrm{b}}(R,z)=\frac{-G\,M_{\mathrm{b}}}{\sqrt{R^{2}+z^{2}}+a_{\mathrm{b}}}\,,
\end{equation}
\noindent
where two free parameters, $M_{\mathrm{b}}$ and $a_{\mathrm{b}}$, are  the total mass and the scale radius of the bulge, respectively.

Finally,  we consider a spherical dark halo, whose potential is modelled with a logarithmic potential in the form \citep[e.g.][]{Binney_Tremaine2008}:
\begin{equation}
\label{eq:Phi_halo}
\Phi_{\mathrm{h}}(R,z)=\frac{v_{\mathrm{h}}^{2}}{2}\,\ln \left(R^{2}+z^{2}+r_{\mathrm{h}}^{2}\right)\,,
\end{equation}
\noindent
where $r_{\mathrm{h}}$ is the core radius and $v_{\mathrm{h}}$ is the circular velocity at large $R$ (i.e., relative to the core radius), respectively. 

The analytical rotation curve is then calculated using the expression. 
\begin{equation}
\label{eq:rot-curve}
V_{rot}(R)=\sqrt{R\times\frac{\partial\Phi_0}{\partial R}\vert_{z=0}}\,,
\end{equation}
\noindent
where the axisymmetric potential $\Phi_0$ is just the sum of the potentials of the Galactic components described above, that is:
\begin{equation}
\label{eq:axisymmetric}
\Phi_0(R,z)=\Phi_\mathrm{thin}+\Phi_\mathrm{thick}+\Phi_\mathrm{HI}+\Phi_\mathrm{H_2}+\Phi_{\mathrm{b}}+\Phi_{\mathrm{h}}\,.
\end{equation}

The physical and structural parameters of the components of the Galactic axisymmetric potential are obtained by applying a fitting procedure, which adjusts the analytical rotation curve Eq.\,(\ref{eq:rot-curve}) to the observed one. For the observation-based rotation curve, we use data of \textrm{H}\,{\scriptsize I}-line tangential directions from \cite{Burton_Gordon1978} and \cite{Fich_Blitz_Stark1989}, CO-line tangential directions from \cite{Clemens1985}, and maser sources data associated with high-mass star-forming regions from \cite{Reid_etal2019} and \cite{Rastorguev_etal2017}. From these data, the Galactic radii and rotation velocities were calculated taking the distance of the Sun from the Galactic center as $R_{0}=8.122$\,kpc \citep{Gravity_collab} and the Local Standard of Rest (LSR) velocity at the Sun $V_{0}=233.4$\,km\,s$^{-1}$ \citep{Drimmel_Poggio2018}. As additional constraints to the fitting procedure, we used the value of the local angular velocity $\Omega_0$, the local disk surface density $\Sigma_0$, and the local surface density within $|z|\leq 1.1$\,kpc. For details of the data used and the fitting procedure, see \cite{Barros2016, Barros2021}. 

The parameters obtained are shown in Table\,\ref{tab-init} and Table\,\ref{tab-init-2}. The contributions of each component of the axisymmetric potential to the Galactic rotation curve are plotted in Fig.\,\ref{fig:rot-curve} (top panel), together with the calculated rotation curve (red line) and the data points. {Figure 1 (bottom panel) also shows the radial profile of the $K_z$ vertical force resulting from the Galactic model calculated at $z=1.1$\,kpc. We note a good agreement between the modeled force and the measured $K_z$--force, based on measurements by \cite{Bovy_Rix2013} (black dots with error bars).}

The stellar orbits  are calculated through the numerical integrations of the equations of motions defined by the Hamiltonian function given as
\begin{equation}
\label{eq:hamiltonian}
H(R,p_r,z,V_z)=\frac{1}{2}\left[p_r^2+\frac{L_z^2}{R^2}+V_z^2\right]+\Phi_0(R,z)\,,
\end{equation}
\noindent
where $p_r$ and $V_z$ are the linear radial momentum and vertical velocity, respectively, while $L_z$ is the angular momentum, which is a constant in the axisymmetric approximation (all momenta are given per mass unit) 


\section{ 3D stellar dynamics on the representative plane}
\label{sec:repres-plane}

\begin{figure}
	\includegraphics[width=0.95\columnwidth]{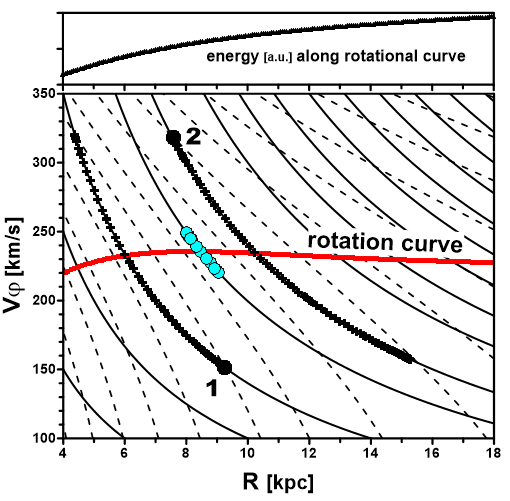}
    \caption{Topology of the Hamiltonian (Eq.\,\ref{eq:hamiltonian}) on the representative $R$--$V_\varphi$  plane. Continuous curves are the levels of the angular momentum $L_z=R\times V_\varphi$, while dashed curves are the energy levels, calculated with the fixed {initial values} $p_R=0$, $z=0$ and $V_z=20$\,km\,s$^{-1}$. Rotation curve is shown by red dots. The projections of the three 3D orbits on the plane are: the Sun's orbit (cyan dots), one orbit starting at the configuration {\bf 1} and other at the configuration {\bf 2} (black dots). Note that the motion occurs always along a fixed $L_z$-value, due to conservation of the angular momentum $L_z$. The conservation of the energy during the motion defines two turning points of each orbit, both lying at the intersections of the corresponding $L_z$-- and energy levels. Top panel: the total energy (see Eq.\,\ref{eq:hamiltonian}), in arbitrary units, calculated along the rotation curve, as a function of $R$.  }
\label{fig:represent-plane-1}
\end{figure}

The dynamical model defined by the Hamiltonian in Eq.\,(\ref{eq:hamiltonian}) is of two-degree-of-freedom. The resulting stellar motion can be formally represented by two coupled oscillations: one, in the radial (equatorial) direction, is described by the pair of the variables $R$--$p_r$, and other, in the vertical direction, described by the pair $z$--$V_z$. The phase space of the system is four-dimensional, that makes it difficult to visualize the stellar orbits. Yet, in this paper, we introduce a representative plane, which allows us to present the main features of the 3D stellar dynamics. This plane is the $R$--$V_\varphi$ plane, where $V_\varphi$ is the tangential velocity defined as $V_\varphi=L_z/R$; we show this plane in Fig.\,\ref{fig:represent-plane-1}. 

To study the stellar motions on the $R$--$V_\varphi$ plane, we fix, in this paper, the initial values of the planar linear momentum at $p_r=0$ and the vertical height at $z=0$. It is worth noting that this choice preserves the generality of the presentation, since all closed stellar motions under the potential given by Eq.\,(\ref{eq:axisymmetric}) pass through these conditions. 
The initial value of the vertical velocity is chosen as $V_z=20$\,km\,s$^{-1}$, except when the dependence of the motion on the initial vertical configuration is analysed (Sect.\,\ref{sec:depend_Vz}). The chosen velocity value is close to the value of the standard deviation of the $V_z$ distribution of the stars from the \textit{Gaia} DR3 \citep{GaiaCollab2016,GaiaCollab2023}, with $|z|<1$\,pc (we obtained $\sigma_{V_{z}}=18$\,km\,s$^{-1}$).

First, we plot on the $R$--$V_\varphi$ plane in Fig.\,\ref{fig:represent-plane-1}  the main  characteristics of the Hamiltonian dynamics: the orbital energy and the angular momentum, which are both constants of motion in our model. The energy levels and $L_z$--levels are shown by dashed and continuous lines, respectively. The rotation curve obtained using Eq.\,(\ref{eq:rot-curve}), is shown by the red curve. 

By definition, the rotation curve is a place of the circular orbits, which are orbits of the minimal energy, for a given $L_z$-value. The geometrical interpretation of this condition is that, given the location of a circular orbit on the rotation curve, the corresponding $L_z$-level is tangent to the minimal energy level on the $R$--$V_\varphi$ plane. This fact can be observed in Fig.\,\ref{fig:represent-plane-1}, considering that the energy is continuously increasing with the increasing radial distances $R$, as shown on the top panel in Fig.\,\ref{fig:represent-plane-1}. The same $L_z$-level  intersects the levels of higher energy always at two points, which are turning points of the radial oscillations at the condition $p_r=0$.

We plot in Fig.\,\ref{fig:represent-plane-1} the direct projections of three 3D stellar orbits: the one is the Sun's orbits (cyan dots) calculated with initial values $R=8.122$\,kpc, $p_r=-12.9$\,km\,s$^{-1}$, $V_\varphi=245.6$\,km\,s$^{-1}$, $z=0.015$\,kpc and $V_z=7.78$\,km\,s$^{-1}$ \citep{Drimmel_Poggio2018}. The orbits of other two fictitious stars (black dots) were starting at points {\bf 1} and {\bf 2}, respectively, with $p_r=0$, $z=0$ and $V_z=20$\,km\,s$^{-1}$. We can verify that all orbits are closed, that is, each one evolves between two turning points, which belong to the same energy levels, along the corresponding $L_z$--levels (constant of motion), around the circular orbits, which lie at the intersection of the corresponding energy and $L_z$--levels.

The detailed analysis of the main features of the stellar motions on the representative $R$--$V_\varphi$ plane can be done in two steps: first, focusing on the radial oscillations on the Galactic equatorial plane, and, then, on the vertical oscillations around the equatorial plane. By understanding the behaviour of each component of the stellar motion in the axisymmetric potential, we can consider new effects produced by their interaction. We will do it in the next sections.

\begin{figure}
	\includegraphics[width=0.99\columnwidth]{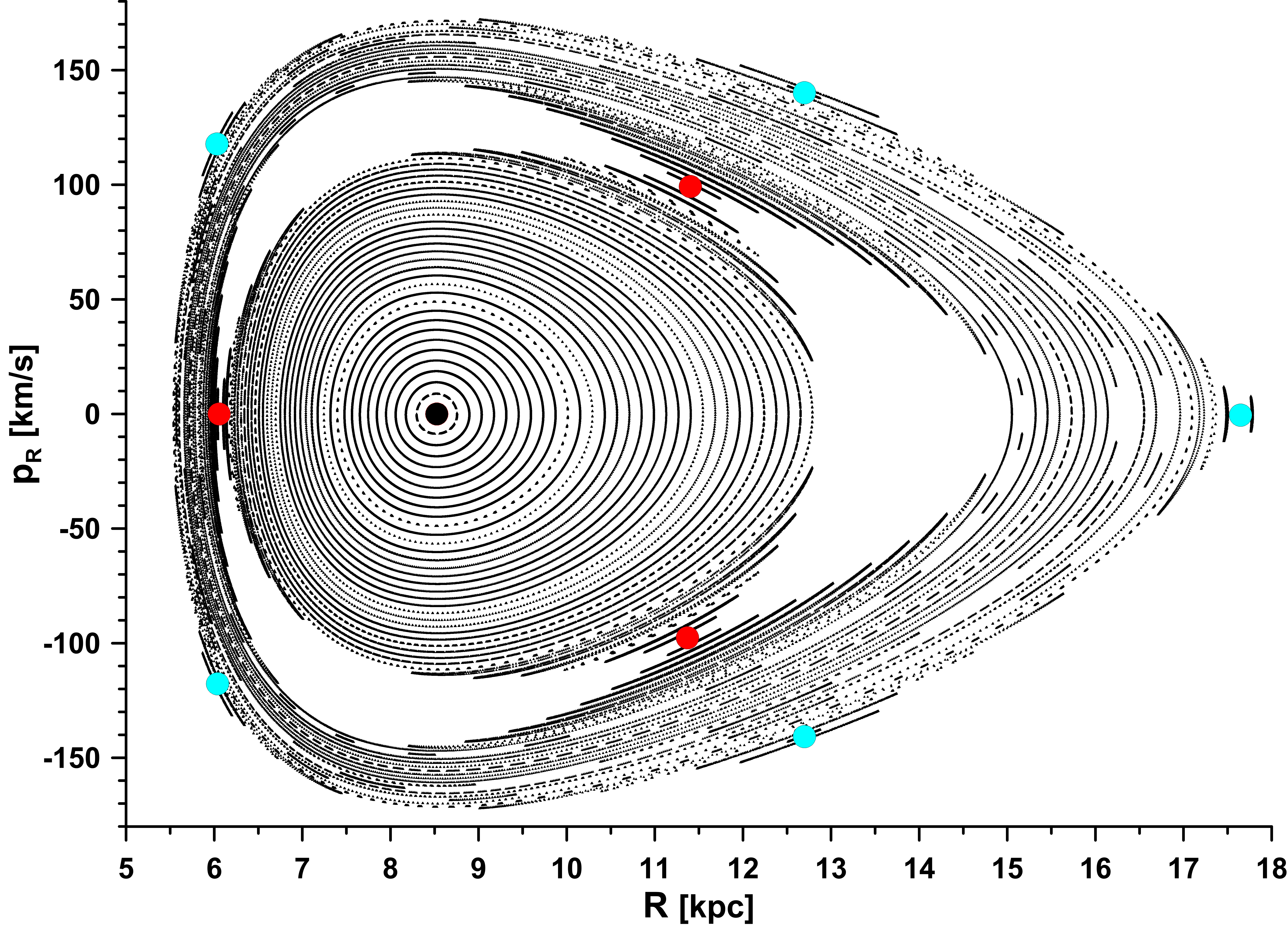}
    \caption{Poincaré map of the stellar 3D orbits calculated with the initial conditions chosen along the Sun's $L_z$--level. The initial vertical conditions were chosen at $z=0$ and $V_z=20$\,km\,s$^{-1}$. The projections are calculated at instants when the star crosses the equatorial plane with $V_z>0$. The  orbits are concentric, with the circular orbit at the center (black dot). This orbit lies at the intersection of the rotation curve and the Sun's $L_z$--level on the $R$--$V_\varphi$ plane, at $R_c=8.522$\,kpc and $V_\varphi=234$\,km\,s$^{-1}$.  Red and cyan dots show the projections of  two periodic orbits on the Poincaré map corresponding to the 2/1 and 4/1 resonances, respectively (see later Fig.\,\ref{fig:ressonant-orbits}).  }
\label{fig:radial-orbits}
\end{figure}


\subsection{Radial oscillations}
\label{sec:radial}

The $R$--$V_\varphi$ plane shown in Fig.\,\ref{fig:represent-plane-1} is commonly used for understanding the radial (or planar) motion of stars. Indeed, due to the conservation of the angular momentum during the motion, the projections of the 3D orbits are aligned along the corresponding $L_z$--levels (continuous lines), as observed in the case of our three examples. Since the orbital energy is also conserved in our problem, the orbit's projections must match the corresponding energy level at conditions $p_r=0$, which was chosen in the construction of the plane. By definition, this condition defines two turning points of a closed orbit, which delimit the path of the radial oscillation on the $R$--$V_\varphi$ plane and, consequently, the maximal ($R_\mathrm{max}$) and minimal ($R_\mathrm{min}$) values of the radial orbital variation (and the tangential velocity $V_\varphi$, for a given $L_z$-value). 

Any $L_z$--level crosses the rotation curve (red curve) at one point at $R_c$ on the $R$--$V_\varphi$ plane in Fig.\,\ref{fig:represent-plane-1}, which corresponds to the circular orbit of the minimal energy for the corresponding $L_z$. At this condition, two turning points merge at one point, characterizing the circular orbit with the zero-amplitude $R$--oscillation, that is, the periodic orbit of the two-degree-of-freedom problem. All other orbits calculated along the same $L_z$--level, are quasi-periodic orbits oscillating with the non-zero $R$--amplitudes, in such a way that larger the deviations from the rotation curve,  larger are the amplitudes of the radial variations. 

The  amplitude of the radial oscillation is related to the planar eccentricity of the orbit as
$$ e=\frac{R_\mathrm{max}-R_\mathrm{min}}{2R_c}\, ,$$
which is uniquely defined by the values of the orbital energy and angular momentum of the stellar orbit. In the conservative problem considered in this paper, it is a constant of motion. 
The gain/loss of the orbital energy due to some external physical processes, without changing the angular momentum of the system, will increase/decrease the amplitude of $R$-oscillation and, consequently, the planar eccentricity. This effect, known as `heating' in the stellar dynamics, is analogous to the tidal gravitational interactions in the planet-satellite systems.

The equatorial motions of the stars are shown  in Fig.\ref{fig:radial-orbits}, where we plot the orbits in the subspace  $R$--$p_R$.
The orbits were calculated through numerical integrations of the equations of motion defined by the Hamiltonian (Eq.\,\ref{eq:hamiltonian}), with initial conditions chosen along the level  $L_z=1995$\,kpc\,km\,s$^{-1}$, which corresponds to the Sun's orbit (see Fig.\,\ref{fig:represent-plane-1}). To obtain the projection of the 3D motion on the Galactic plane, we gather the orbital coordinates at the instants when the orbits cross the equatorial plane (the condition $z=0$), in the direction of the positive $z$-values (the condition $V_z>0$). This approach is known as the Poincaré surfaces of section method, which allows us to plot the radial motion separately from the vertical one. 
\begin{figure}
	\includegraphics[width=1.0\columnwidth]{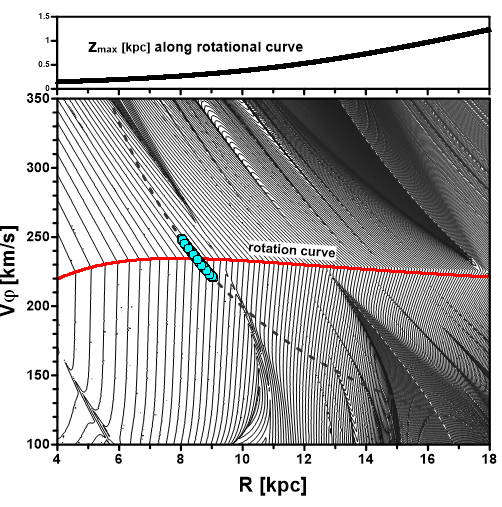}
    \caption{Representative $R$--$V_\varphi$ plane shown in Fig.\,\ref{fig:represent-plane-1}, with the same rotation curve (red dots) and the Sun's orbit (cyan dots) along the $L_z$--level (dashed line).  The equidistant levels are used to represent the amplitude of the vertical oscillation of the stellar orbits (see top panel to associate the absolute $z_\mathrm{max}$-values), with the initial conditions $z=0$ and $V_z=20$\,km\,s$^{-1}$. Top panel: the amplitude of the vertical oscillation, $z_\mathrm{max}$, calculated along the rotation curve, as a function of $R$. 
    }
\label{fig:represent-plane-2}
\end{figure}

Figure \ref{fig:radial-orbits} shows that all orbits in the $R$--$p_R$ subspace oscillate around a circular periodic orbit (black dot), which lies at the intersection of the corresponding $L_z$--level and the rotation curve,  at $R_c=8.522$\,kpc and $V_\varphi=234$\,km\,s$^{-1}$ on the $R$--$V_\varphi$ plane in Fig.\,\ref{fig:represent-plane-1}. This periodic orbit appears as a fixed point on the $R$--$p_R$ plane, indicating that the amplitude of the radial oscillation is zero (or $e=0$). At the same time, the vertical motion started with the initial $V_z=20$\,km\,s$^{-1}$ has non-zero amplitude; it is a simple oscillation around the equatorial plane ($z=0$).  

In the vicinity of the fixed point (black dot), between 7.0\,kpc and 10.5\,kpc, the concentric low-eccentricity orbits, with $e < 0.2$, evolve in good agreement with the epicyclic approximation \citep[e.g.][]{Binney_Tremaine2008}. 
{The orbit of the Sun, with eccentricity of 0.061, is located in this region of the phase space. The values of the radial and vertical frequencies for the Sun's $L_z$ level, derived under the epicyclic approximation, are equal to 0.00542 and 0.01387  (in units of 1/Myr), respectively, that corresponds to the radial and vertical periods of 184.5\,Myr and 72.07\,Myr. These periods derived from the power spectrum of the Sun's orbit, calculated numerically through our model, are of 161.3\,Myr and 75.2\,Myr, respectively.}

The orbits with the increasing eccentricity, however, deviate from the epicyclic approximation and, when the eccentricity approaches 0.5, a notable feature appears in Fig.\ref{fig:radial-orbits}. Indeed, at the radial distances  beyond 12\,kpc, we can clearly observe a lack of circulating orbits, but there are some islands on the Poincaré map. 
{As shown below, the orbits in this region are different structurally from the low-eccentricity orbits and are related to the new kind of motion  mode.}  
The planar axisymmetric model is not able to explain this feature, thus, we proceed investigating the vertical motion and its interaction with the planar component.

\begin{figure}
	\includegraphics[width=0.99\columnwidth]{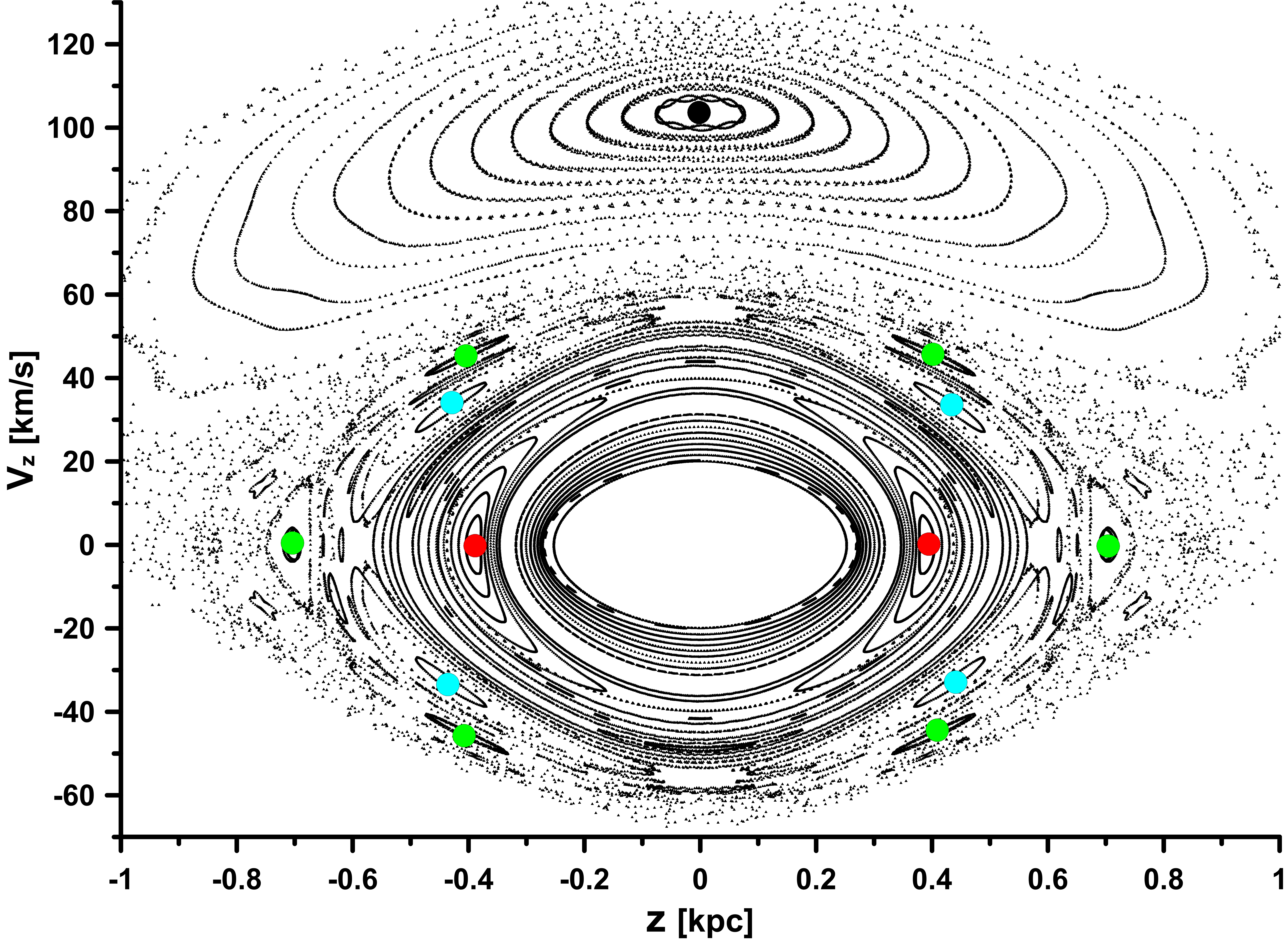}
    \caption{Same as in Fig.\,\ref{fig:radial-orbits}, except the projections were calculated at conditions $p_R=0$ and $\dot{p}_R>0$ on the $z$--$V_z$ plane.   Red, cyan and green dots show the projections of three periodic orbits on the Poincaré map corresponding to the 2/1, 4/1 and 6/1 resonances, respectively.  The large island of the orbits oscillating around the center at $z=0$\,kpc and $V_z=103.0$\,km\,s$^{-1}$ (black dot) is related to the strong 1/1 resonance (see later Fig.\,\ref{fig:ressonant-orbits}).
    }
\label{fig:vertical-orbits}
\end{figure}

\subsection{The vertical oscillations} 
\label{sec:vertical}

The same representative $R$--$V_\varphi$ plane shown in Fig.\,\ref{fig:represent-plane-1} can be used to study the stellar motion in the vertical direction. 
For this, we plot in Fig.\,\ref{fig:represent-plane-2},  by continuous lines, the levels of the maximal altitude over the equatorial plane that a star reaches during its vertical oscillation ($z_\mathrm{max}$). To obtain $z_\mathrm{max}$, we integrate numerically, over several billions years, the equations of motion defined by the Hamiltonian (Eq.\,\ref{eq:hamiltonian}), over the 200$\times$200 grid of the initial conditions covering the $R$--$V_\varphi$ plane. All orbits started with $p_R=0$ on the galactic equatorial plane (at $z=0$) and with the same initial vertical velocity  $V_z=20$\,km\,s$^{-1}$. The panel on the top of Fig.\,\ref{fig:represent-plane-2} shows $z_\mathrm{max}$ obtained in this way for the circular orbits located along the rotation curve (red curve). We plot also the projection of the Sun's orbit (cyan dots) along the $L_z$--level (dashed line) on the $R$--$V_\varphi$ plane. 

To understand the $z_\mathrm{max}$--evolution on the Galactic $R$--$V_\varphi$--plane, we consider first the nearly circular orbits, distributed along the rotation curve. For these low-eccentricity orbits, the amplitude of the vertical oscillation is smoothly increasing with the growing Galactocentric distance, as shown on the top panel in Fig.\,\ref{fig:represent-plane-2}. This result is consistent with the radially decaying exponential mass distribution model applied and is in agreements with the 
observed increase of the scale height with the increase of the Galactic radius of the disk stars, which is the well-known flaring of the Galactic disk \citep{LopezCorredoira2002,Amores2017,Robin2022}.

However, the continuous evolution of the $z_\mathrm{max}$--levels is interrupted in the regions of the increasing planar eccentricities when we move away from the rotation curve on the $R$--$V_\varphi$ plane in Fig.\,\ref{fig:represent-plane-2}, in the direction of both the higher and lower tangential velocities. To understand this behaviour, it is worth noting that, in domains of regular oscillations, small changes in the initial conditions lead to small changes of the elements of the orbits, in particular, the amplitude $z_\mathrm{max}$. Consequently, the levels  suffer slight displacements on the map when the initial conditions are gradually changed. However, in the vicinity of the resonances, small changes in the initial configurations produce large changes of the orbital elements, forming singular structures, such as resonant islands and stochastic layers. On the dynamical map in Fig.\,\ref{fig:represent-plane-2}, these structures appear in the form of stalactites of different widths.  

\begin{figure}
	\includegraphics[width=0.9\columnwidth]{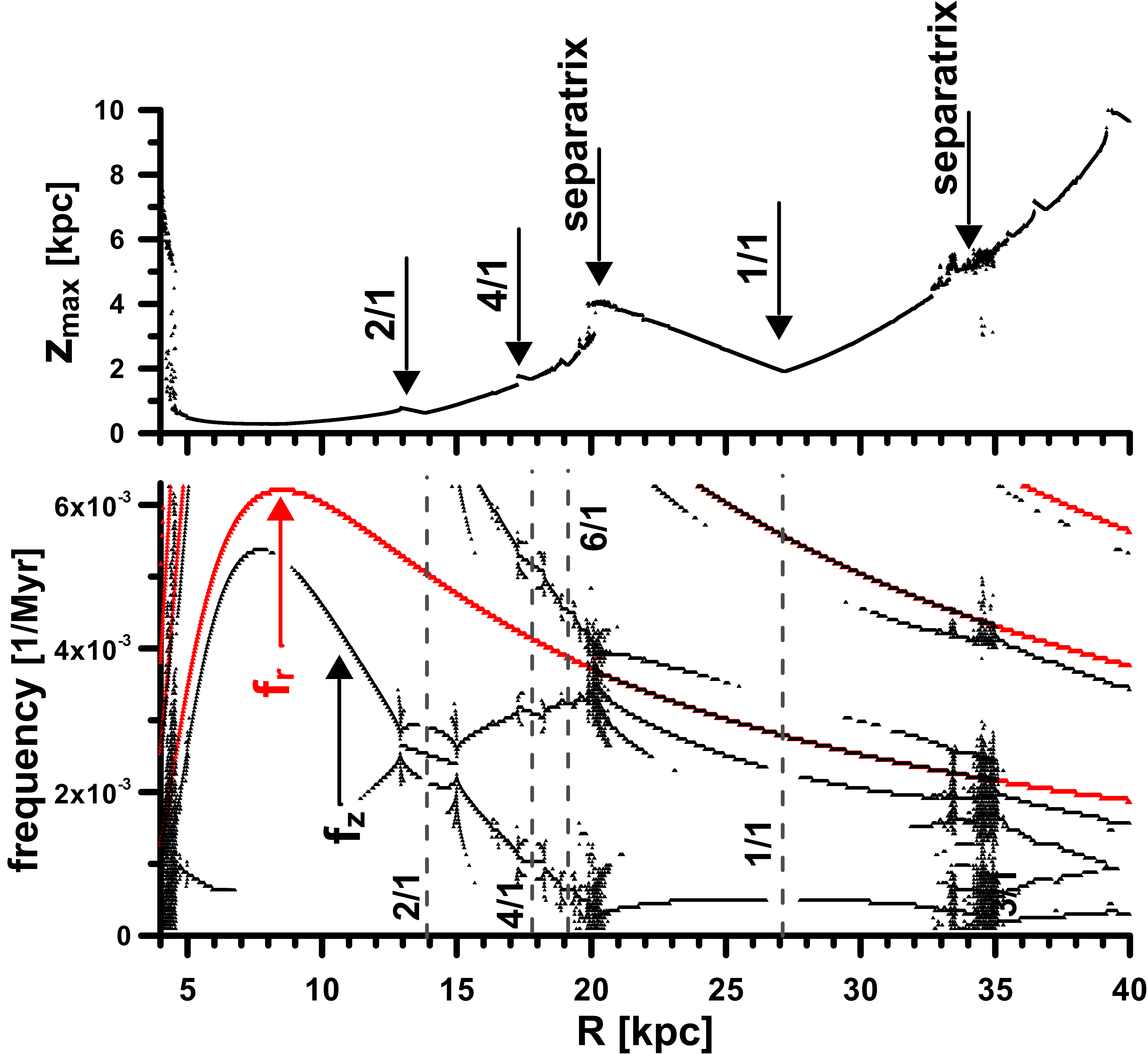}
    \caption{Top: Amplitude of the vertical oscillation of stars as {a function of the initial Galactocentric distance. The initial values of $V_\varphi$ are chosen along the Sun's $L_z$--level, while $z=0$ and $V_Z=20$\,km\,s$^{-1}$}. Bottom: Dynamical spectrum of the stellar orbits calculated along the Sun's $L_z$--level. The nominal positions of some resonances are shown by vertical dashed lines.
    }
\label{fig:spectr}
\end{figure}

On the Poincaré map in Fig.\,\ref{fig:vertical-orbits}, the same structures appear as chains of islands contrasting with the orbits, which regularly circulate around the origin. To construct this map, we pick up the orbital coordinates $z$ and $V_z$ at the instants when the star is in the turning point of its orbit ($p_r=0$) of the minimal radial distance ($\dot{p}_r > 0$). 
{The initial configurations of the orbits were chosen along the Sun's $L_z$--level and $z$ and $V_z$ fixed at 0 and 20\,km\,s$^{-1}$, respectively, as in Fig.\,\ref{fig:radial-orbits}. However, in this case, the initial $R$--values were extended  to the very large radial distances, up to 27\,kpc.}

{Comparing this map to the map of the equatorial motion shown in Fig.\,\ref{fig:radial-orbits}, we verify that the behaviour in the vertical direction is more complicated. We can observe the projections of the qualitatively different types: there are circulating orbits of regular motion at smaller stellar altitudes. With the growing height, the chains of islands of different width appear; they are separated by the stochastic layers. A large island dominates the region of  high vertical velocities in the positive half-plane of the map and is surrounded by the sea of chaotic motion.    
To identify the cause of this behaviour, we  apply a special spectral analysis method and describe the results obtained in the next section.}

\begin{figure*}
\begin{center}
	\includegraphics[width=0.9\columnwidth]{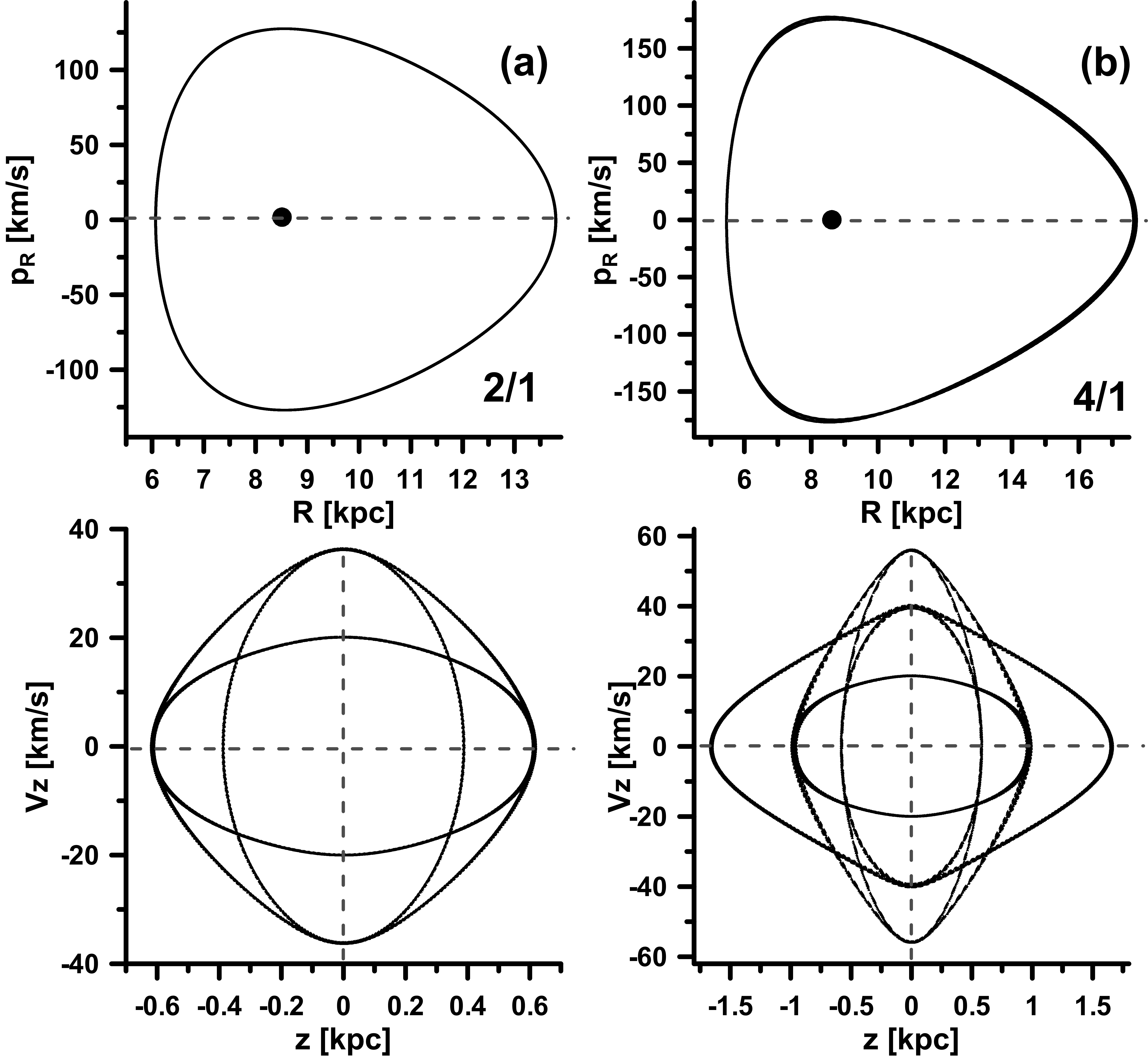}
	\includegraphics[width=0.9\columnwidth]{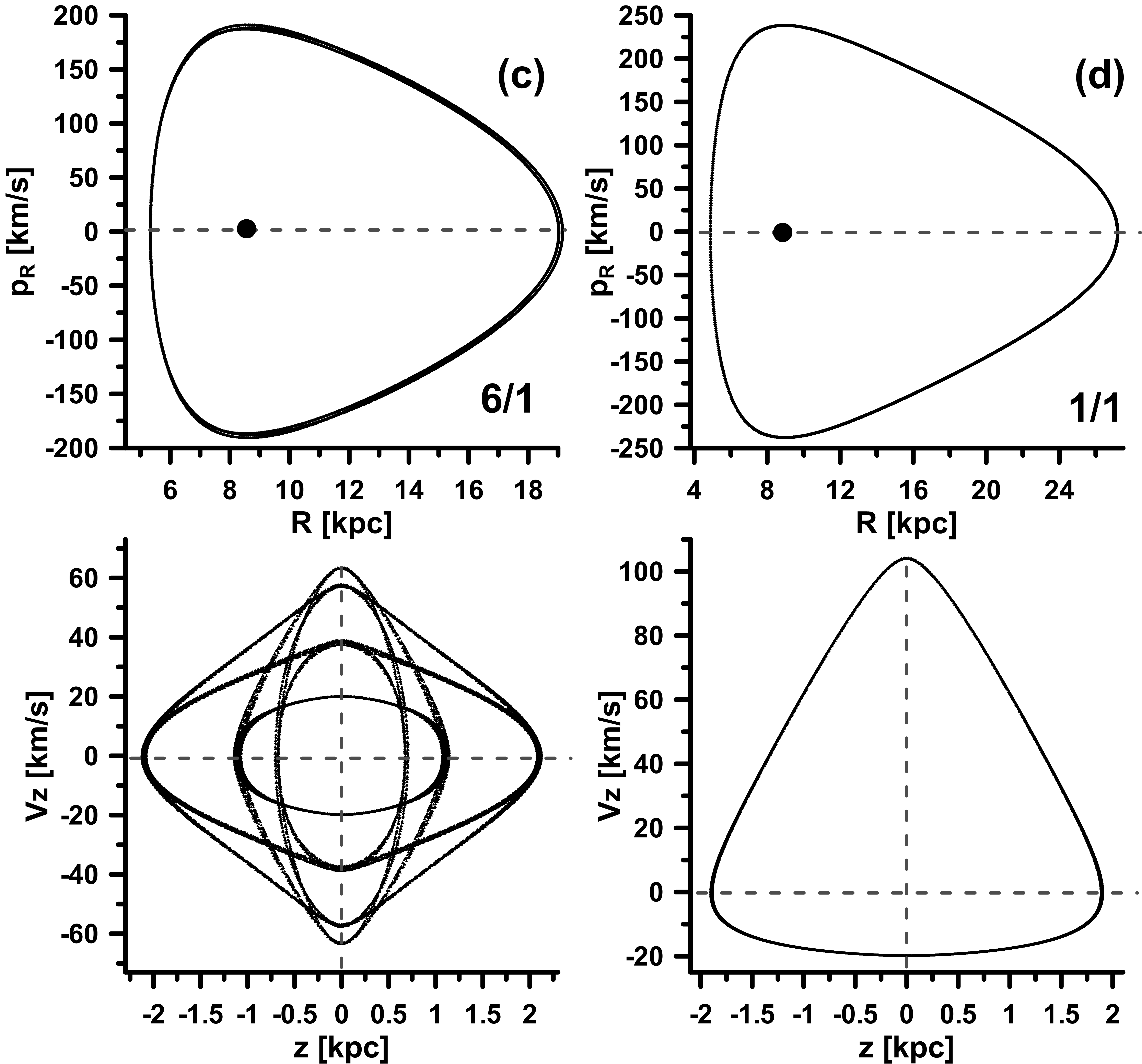}
    \caption{Resonant periodic orbits on the $R$--$p_R$ plane (top row) and $z$--$V_z$ plane (bottom row). The initial conditions were chosen along the Sun's $L_z$--level, while $z=0$ and $V_z=20$\,km\,s$^{-1}$. Column (a): the 2/1 resonant orbit with $e=0.45$ (red dots on the Poincaré maps in Figs.\,\ref{fig:radial-orbits} and \ref{fig:vertical-orbits}). Column (b): the 4/1 resonant orbit with $e=0.72$ (cyan dots in Figs.\,\ref{fig:radial-orbits} and \ref{fig:vertical-orbits}).  Column (c): the 6/1 resonant orbit with $e=1.01$ (green dots in Fig.\,\ref{fig:vertical-orbits}).  Column (d): the 1/1 resonant orbit with $e=0.802$ (black dot in Fig.\,\ref{fig:vertical-orbits}).
    }
\label{fig:ressonant-orbits}
\end{center}
\end{figure*}


\section{The \lowercase{\textit{\large e}}--\lowercase{\textit{\large z}} resonances}
\label{sec:e-z-resonances}

To understand the features of the resonant motion, we start plotting the amplitude of the vertical oscillation ($z_\mathrm{max}$) as a function of the initial radial distance $R$ on the top panel in Fig.\,\ref{fig:spectr}. In contrast to one another shown on the top panel in Fig.\,\ref{fig:represent-plane-2}, the amplitude  $z_\mathrm{max}$ was now calculated with initial conditions chosen along the $L_z$--level corresponding to the Sun's orbit. In this case, all orbits are eccentric, with exception of one located at $R_c=8.522$\,kpc, at the intersection of the $L_z$--level  with the rotation curve in Fig.\,\ref{fig:represent-plane-2}. 

For the orbits starting along the same $L_z$--level at the same initial configurations in the vertical direction ($z=0$ and $V_z=20$\,km\,s$^{-1}$), the minimal value of $z_\mathrm{max}$ is associated with the circular orbit, at $R_c=8.522$\,kpc.  The vertical amplitude is generally increasing when the orbital eccentricity grows with both increasing and decreasing distances from the circular orbit. However, this evolution of $z_\mathrm{max}$ is not monotonous, but shows sudden increase/decrease at some radial distances. The nature of this behaviour can be analyzed using the dynamical power spectrum method, which allows us to detect the change of a regime of motion through the analysis of the evolution of the main frequencies of the dynamical system {\citep[detailed description of the method and its applications to different systems are found in e.g.][]{2002Icar..158..343M,Michtchenko2017}}.

Figure \ref{fig:spectr}\,bottom shows two proper frequencies, $f_r$ and $f_z$, which are the frequencies of the radial and vertical oscillations, respectively, as functions of the radial distances $R$. The evolution of the $f_r$--frequency (red dots) and its harmonics is monotonous over the whole $R$--range; it reaches the maximal value at $R_c=8.522$\,kpc, that is expected since the circular orbit is an equilibrium solution of the effective potential
$\Phi_\mathrm{eff}=\frac{L_z^2}{2R^2}+\Phi_0(R,z)$, with the constant $L_z$. On the other hand, the behaviour of the $f_z$--frequency (black dots) on the dynamical spectrum is highly non-harmonic and we can observe the  appearance of bifurcations,  stable islands and an erratic scattering of the dots, which is characteristic of the chaotic motion. 

Analyzing the evolution of the proper frequencies in Fig.\,\ref{fig:spectr}\,bottom, we verify that bifurcations occur when  the beats between the two independent frequencies occur. This condition is known as resonances between distinct modes of motion, which lead to changes in the regime of motion. {One beat occurs around $R=14$\,kpc, where $f_R\cong 2\,f_z$, giving origin to the island of the 2/1 resonant motion. The periodic orbit of the 2/1 resonance is shown in the subspaces $R$--$p_R$ (top) and $z$--$V_z$ (bottom) in Fig.\,\ref{fig:ressonant-orbits}\,(column {\bf a}); the projections of this orbit in the Poincaré maps are  shown by red dots in Figs.\,\ref{fig:radial-orbits} and \ref{fig:vertical-orbits}}.

{The beating frequencies are also observed at $R=17.8$\,kpc in the dynamical spectrum in Fig.\,\ref{fig:spectr}\,bottom; this event is associated to the 4/1 resonance. The radial and vertical oscillation modes of the corresponding periodic orbit are shown in Fig.\,\ref{fig:ressonant-orbits}\,(column {\bf b}); the projection of this orbit in the Poincaré maps is  shown by cyan dots in Figs.\,\ref{fig:radial-orbits} and \ref{fig:vertical-orbits}. In Fig.\,\ref{fig:vertical-orbits}, we also show, by green dots, the projection of the 6/1 resonant orbit (see Fig.\,\ref{fig:ressonant-orbits}\,\textbf{c}); the location of this resonance in the dynamical spectrum  is shown at $R=19.2$\,kpc.}

We denote these resonances as the $e$--$z$ resonances, where $e$ and $z$ denote planar eccentricities and vertical heights of stellar orbits, respectively. The most important from the $e$--$z$ resonances is the 1/1 resonance, whose domain is large and separated by the layers of chaotic motion (separatrices). For a given value of $L_z$, it appears at very large  Galactic distances; in the dynamical map in Fig.\,\ref{fig:spectr}, its domain is extended from 20\,kpc to 35\,kpc. The motion inside the 1/1 resonance is stable, with the amplitude of the vertical oscillation reaching its minimal  value at the center (locus) of the resonance, at $\sim$27\,kpc; this locus appears as a fixed point (black dot) in the Poincaré map in Fig.\,\ref{fig:vertical-orbits}. The locus is a periodic orbit of the 1/1 resonance shown in Fig.\,\ref{fig:ressonant-orbits}\,(column \textbf{d}). The neighborhood of the 1/1 resonance is generally a domain of highly unstable motion. This is due to the overlap of the high-order resonances of the type $f_r/f_z \cong m/n$ ($m$ and $n$ are integers). 

Finally, it is worth emphasizing that, comparing the evolution of the two frequencies in Fig.\,\ref{fig:spectr}\,bottom, we note the  qualitative difference in the behaviour of two independent modes: while the radial motion seems to be unaffected by the passages through the resonances, their impact on the vertical motion is significant. This could be explained by  the peculiar characteristics of the massive disk potential.



\section{The dependence on the initial vertical velocity $V_z$}
\label{sec:depend_Vz}

\begin{figure}
	\includegraphics[width=1.0\columnwidth]{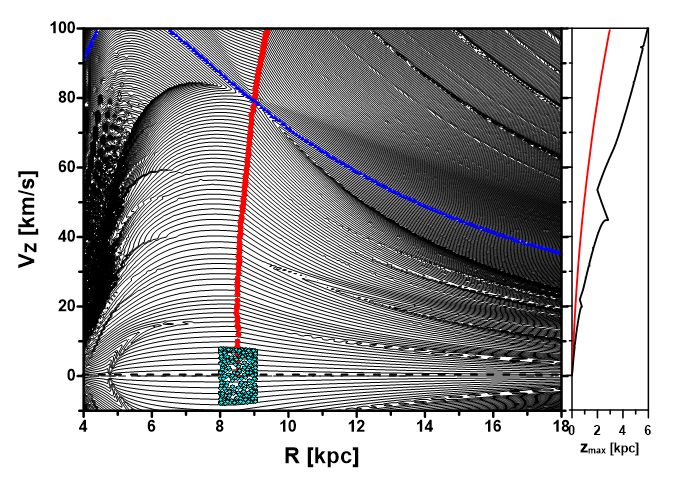}
    \caption{The $R$--$V_z$ plane of initial conditions chosen along the Sun's $L_z$-level, with $p_R=0$ and $z=0$. The equidistant levels of the maximal vertical deviation from the equatorial plane, $z_\textrm{max}$, are shown by continuous curves; its value is mainly increasing with the increasing $V_z$  (see beside panel to associate the absolute $z_\mathrm{max}$-values). The red curve shows the location of the circular orbits with $L_z=1995$\,kpc\,km\,s$^{-1}$ as a function of the initial $V_z$. The structures, which are characteristic of the $e$--$z$ resonances, appears on both sides of the red curve, expanding  their domains with increasing eccentricities. The blue curve shows loci of the very strong 1/1 resonance. The projection of the Sun's orbit is shown by cyan dots; its maximal $z$--amplitude is around 0.85\,kpc. Beside panel: $z_\mathrm{max}$-values as a function of the initial $V_z$, for circular orbits (red curve) and orbits with the initial $R=13$\,kpc (black curve).
    }
\label{fig:plane-R-Vz}
\end{figure}

In this section, we investigate the resonant behaviour as a function of the initial vertical velocity, $V_z$. For this, we introduce  the representative plane $R$--$V_z$ of the initial conditions and fix the rest of the variables at $p_r=0$,  $z=0$ and $L_z=1995$\,kpc\,km\,s$^{-1}$, this last corresponding to the Sun's angular momentum.

Figure \ref{fig:plane-R-Vz} presents the amplitude of the vertical oscillation in the form of the $z_\textrm{max}$--levels on the representative $R$--$V_z$ plane. The red curve shows the positions of the circular orbits {of the constant $L_z$}, which are slightly dislocated in the direction of the larger Galactocentric distances, when the initial vertical velocities increase. The panel beside the $R$--$V_z$ plane allows us to quantify the $z_\textrm{max}$--amplitude, showing its values calculated along the loci of the circular orbits (red curve). The smoothed evolution of this quantity with the increasing initial $V_z$ can be observed.

The bifurcation structures, which are characteristic of the resonances, appear outside the loci of the circular orbits in Fig.\,\ref{fig:plane-R-Vz} and are strengthened with the increasing planar eccentricities of the orbits. The black curve on the beside panel shows the evolution of the  $z_\textrm{max}$--amplitude calculated along the constant initial $R=13$\,kpc. We can observe the behaviour which is similar to that shown on the top panel in Fig.\,\ref{fig:spectr}, indicating the passages through some $e$--$z$ resonances. The most prominent passage is through the 1/1 resonance, whose loci are shown by the blue dots in Fig.\,\ref{fig:plane-R-Vz}. The projection of the Sun's trajectory on the  $R$--$V_z$ plane is shown by the cyan dots. Its  proximity  to the rotation curve avoids the capture in one of the resonances. 


\section{Conclusions}

In this paper, we report the existence of the resonant motion of the kind $e$--$z$ ($e$ and $z$ being the planar eccentricity and  the vertical altitude of stellar orbits, respectively), produced by the axisymmetric Galactic 3D potential. 

We investigate the spacial motion of the stars applying the elaborated model for the axisymmetric potential of the Galactic disk \citep{Barros2016}. The model accounts for the combined gravitational effects due to the two stellar disks (thin and thick disks) and two gaseous disks (\textrm{H}\,{\scriptsize I} and $\mathrm{H}_2$ disks). Moreover, to account for a radially exponential Galactic mass distribution, each of the disks is approximated by the composition  of  the three Miyamoto-Nagai disk proﬁles \citep{Smith2015}.  The physical and structural parameters of the modeled components were chosen to correspond to the observable rotation curve of the Galaxy. 

The motion of stars in the axisymmetric Galactic potential is investigated in the whole phase space applying several techniques from the Celestial Mechanics. The one of those is the dynamical mapping of the representative plane chosen here as the $R$--$V_\varphi$ plane. Based on the conservation laws, we identify two independent modes of the stellar motion, radial and vertical ones.  The radial motion is an oscillation around the circular solution, which is a circular orbit belonging to the rotation curve and characterized by the same angular momentum of the radial mode. The vertical motion is an oscillation around the equatorial Galactic plane. 

For nearly circular orbits, with small planar eccentricities, two oscillations are weakly coupled. It is worth noting that the circular orbits survive even in the case when vertical velocities are very high. However, when the planar eccentricities increase, the coupling between two modes of motion also increase. Their interaction is better visualised in the dynamical power spectra, which clearly show the regions in the phase space where two proper frequencies are commensurable. This beating frequencies indicate the resonant zones, which we denote as $e$--$z$ resonances. 

In the Hamiltonian theories, resonances are fundamental properties of dynamical systems with two or more degrees of freedom. They occur in domains of the phase space where the frequencies of the independent modes of motion are commensurable. The locations and the sizes of the resonant domains are strongly dependent on the  physical parameters of the model adopted to describe the system under study. Thus, associating some observable structures of the objects to the resonances, we can assess the reliable ranges of the parameters of the applied model. One example of this approach is shown in \cite{Michtchenko2018}, where we relate the moving groups in the solar neighbourhood to the Lindblad resonances produced by the spirals perturbations, and estimate the spiral strength and pattern speed.

The main features of the $e$--$z$ resonant motion are sudden increase/decrease in the amplitude of the vertical oscillation,  capture and trap of the stars inside the stable resonant zones, enhancing the density, and deplete of the regions close to the separatrices of the resonances. This behaviour is characteristic of the vertical oscillations, while the planar radial motion is only slightly affected by the $e$--$z$ resonances; this is probably due to peculiar  properties of a disk potential at all. 

It is worth emphasizing that, despite the fact that the resonances occur at very large Galactic distances, the high eccentricities of the resonant orbits allow us their detection in the solar neighbourhood, according to Fig.\,\ref{fig:ressonant-orbits}. Indeed, analyzing the behaviour of the \textit{Gaia}-eDR3 sample from the solar vicinity ($R_{\odot}\pm0.2$\,kpc and $|p_R|\geq 100$\,km\,s$^{-1}$), we have detected 10\% (from one thousand randomly chosen stars) of the objects showing the resonant oscillations, mainly, inside the 2/1 resonance. Of course, whether these stars are really evolving inside the $e$--$z$ resonances, it depends strongly on the parameters adopted in our model.  Thus, we should be able to observe  the described manifestations of the $e$--$z$ resonances analysing the distributions of the observable proper elements of the stars. 
The comparison to theoretical predictions could allow us to improve unknown values of the parameters, which describe the observable Galactic mass distribution.

\begin{acknowledgements}

We acknowledge the anonymous referee for the detailed review and for the many helpful suggestions which allowed us to improve the manuscript. This work was supported by the Brazilian CNPq, FAPESP, and CAPES. This work has made use of the facilities of the Laboratory of Astroinformatics (IAG/USP, NAT/Unicsul), funded by FAPESP (grant 2009/54006-4) and INCT-A. This work has made use of data from the European Space Agency (ESA) mission {\it Gaia} (\url{https://www.cosmos.esa.int/gaia}), processed by the {\it Gaia} Data Processing and Analysis Consortium (DPAC, \url{https://www.cosmos.esa.int/web/gaia/dpac/consortium}). Funding for the DPAC has been provided by national institutions, in particular the institutions participating in the {\it Gaia} Multilateral Agreement.

\end{acknowledgements}

\bibliographystyle{aa} 
\bibliography{refs} 

\end{document}